\documentstyle[11pt]{article}

\def\fii{\varphi}
\def\al{\alpha}
\def\be{\beta}
\def\ga{\gamma}
\def\ro{\varrho}
\def\si{\sigma}
\def\eps{\varepsilon}
\def\la{\lambda}
\def\lap{\bigtriangleup\,}
\def\te{\vartheta}
\def\B{\Bigl}
\def\=d{\,{\buildrel\rm def\over =}\,}
\def\sqr#1#2{{\vcenter{\vbox{\hrule height.#2pt\hbox{\vrule width.#2pt height#1pt \kern#1pt \vrule width.#2pt}\hrule height.#2pt}}}}
\def\sq{\hbox{\rlap{$\sqcap$}$\sqcup$}}
\def\w{\mathchoice\sqr45\sqr45\sqr{2.1}3\sqr{1.5}3\,} 
\def\eh{{\scriptstyle{1\over 2}}}
\def\ev{{\scriptstyle{1\over 4}}}
\def\lra{\longleftrightarrow}
\def\d{\partial}
\def\dh{\mathop{\vphantom{\odot}\hbox{$\partial$}}}
\def\dl{\dh^\leftrightarrow}
\def\io{\int{d^3k\over\sqrt{2\omega}}\,}
\def\pdh{(2\pi )^{-3/2}}
\def\Hp{{\cal H_{\rm phys}}}
\def\kon#1#2{\vbox{\halign{##&&##\cr\lower4pt \hbox{$\scriptscriptstyle\vert$}\hrulefill &\hrulefill\lower4pt \hbox{$\scriptscriptstyle\vert$}\cr $#1$&$#2$\cr}}}
\def\sgn{{\rm sgn}}
\newcommand{\comment}[1]{}

\begin{document}

\comment{
\noindent TEST DEFINITIONS \\
Letters \\
$ \fii \al \be \ga \ro \si \eps \la \lap \te \B( \B) $ \\
Symbols \\
$ \=d $ \\
Squares \\
$  \sqr{20}{4} \sq \w $ \\
Fractions \\
$ \eh \ev $ \\
Arrows \\
 $ \lra $ \\
Derivatives \\
$ \d_\mu \dh_\mu \dl_\mu \=d $ \\
Integrals \\
$ \io $ \\
Contractions \\
$ \kon34 \kon{\al}{\be} $ \\
Miscellaneous \\
$ \pdh  \Hp $ \\
}

\title{Dark matter in galaxies according to the tensor-four-scalars theory II }
\author{G\"unter Scharf \footnote{e-mail: scharf@physik.uzh.ch} and Fabrice Larere
\footnote{e-mail: fabrice.larere@gmail.com}
\\ Institut f\"ur Theoretische Physik, 
\\ Universit\"at Z\"urich, 
\\ Winterthurerstr. 190 , CH-8057 Z\"urich, Switzerland}

\date{}

\maketitle\vskip 3cm

\begin{abstract}  We continue to study the spherically symmetric vacuum solutions of the tensor-four-scalars theory which is a modification of general relativity. Our aim is to construct vacuum solutions with asymptotically constant circular velocity curves which can be used to represent the dark halo outside the baryonic matter. These solutions are obtained by means of a new non-standard gauge in the field equations. The metric is expressed in terms of the rotation curve $V(r)$ in closed form.

\end{abstract}
\vskip 1cm
{\bf PACS numbers: 04.06 - m; 04.09 + e}

\newpage

\section{Introduction}
As pointed out in the first part of this series [1] the tensor-four-scalars theory is the classical theory corresponding to the massless limit of massive gravity.
On the theoretical side, taking quantum gauge theories for spin-1 and spin-2 gauge fields as the basic framework to describe nature [2-5], this theory has the same right for being considered as fundamental as Einstein's theory. In this respect our theory differs from the many modifications of general relativity discussed in the literature.
On the experimental and observational side the theory is very interesting because, due to the four massless scalar fields, it has additional degrees of freedom which tentatively might explain the dark matter phenomenology without introducing hypothetical particles.

There is a second important aspect of the dark matter problem: {\it the right choice of the gauge.} Since gravity is described by a spin-2 gauge theory different gauges are possible. In the case of static spherically-symmetric problems described by means of spherical coordinates $t, r, \te, \phi$ the standard gauge defined by $g_{22}\equiv g_{\te\te}=-r^2$ is widely used. The corresponding
vacuum solution in general relativity (disregarding rotation) is the Schwarzschild metric. It is unique due to Birkhoff's theorem. This choice of gauge is appropriate for small bodies up to stars and clusters, but obviously not for galaxies. The reason is that the corresponding circular velocity (rotation curve) must decrease as $V^2\sim 1/r$. Until now no such decrease has been observed in galaxies. Therefore, it is highly desirable to investigate other gauges which lead to solutions with different asymptotic behavior. This is one purpose of this paper, the other being the tensor-four-scalars theory.

The paper is organized as follows. In the next section we start from a discussion of the measurement of the rotation curves in galaxies. This will lead us to the introduction of a new non-standard gauge. In section 3 we study spherically symmetric vacuum solutions in this gauge. We reduce the field equations to a single second order ordinary differential equation in the radial coordinate $r$. In section 4 we integrate this equation and express the metric functions by the circular velocity $V(r)$ in closed form. In the last section we fit the tail of the rotation curve of M33 given by Corbelli [8] to a simple vacuum solution. 

\section{Choosing a non-standard gauge}

Dark matter has been introduced in astronomy to explain the rotation curves of galaxies. Therefore, we must discuss how the circular velocities and the radial distances are measured. We use spherical coordinates $x^\mu =(t,r,\te,\phi)$ throughout which define our laboratory system on earth. The radial distance $r$ is measured in light years or kpc. The distances of nearby galaxies can be determined by measuring the apparent luminosities $l$ of some standard candle (as the classical Cepheids) with a priori known absolute luminosity $L$. Then the so-called luminosity distance is given by
$$d_L=\sqrt{{L\over 4\pi l}}.\eqno(2.1)$$
In the large we assume Euclidean geometry, the space is only locally curved in the vicinity of galaxies. On the other hand we only consider nearby galaxies so that the cosmic expansion can be neglected. The radial distances $r$ within the galaxy measured on earth as angles $\al$ in arcmin are converted to kpc by
$$r=\al d_L{2\pi\over 360\cdot 60}.$$

The rotation velocity of stars or neutral hydrogen gas is measured by means of the Doppler shift of spectral lines
$${\nu_{\rm obs}\over\nu}=1-V_{\rm rel}+O(V^2).\eqno(2.2)$$
Here $\nu$ is the frequency of the light source at rest and $\nu_{\rm obs}$ the frequency actually measured by the observer on earth, $V_{\rm rel}$ is the component of the relative velocity $\vec V$ along the direction from observer to the light source. Following Weinberg ([6], chapter 3/2) we introduce the freely falling coordinate system $\xi^\al$ of the moving star. In this system the star is at rest and the observer on earth moves with the 4-velocity
$$u^\al={d\xi^\alpha\over d\tau}=(u^0,\vec u),\quad u^0=(1-\vec V^2)^{-1/2},\quad \vec u=u^0\vec V,\eqno(2.3)$$
because in the locally inertial coordinates special relativity holds. Here $\tau$ is the proper time
$$d\tau^2=\eta_{\al\beta}d\xi^\al d\xi^\beta,\eqno(2.4)$$
$\eta_{\al\beta}={\rm diag}(1,-1,-1,-1)$ is the Minkowski tensor and the speed of light is put $=1$. The Doppler shift is given by the special relativistic result ([6], eq.(2.2.2))
$${\nu_{\rm obs}\over\nu}={\sqrt{1-\vec V^2}\over 1+V_{\rm rel}},\eqno(2.5)$$
which yields (2.2) in leading order.

In addition Weinberg introduces a laboratory coordinate system $x^\mu$ which in our case is attached to the observers telescope. To simplify the following discussion we assume that the astronomer on earth has corrected his measurements for the motion of the earth with respect to the center of the galaxy, so that we can choose the center of the galaxy as origin of the laboratory coordinate system. Now the star moves on a geodesic ([6], eq.(3.2.3))
$${d^2x^\mu\over d\tau^2}+\Gamma^\mu_{\al\beta}{dx^\al\over d\tau}{d x^\beta\over d\tau}=0\eqno(2.6)$$
where $\Gamma_{\al\beta}^\mu$ are the Christoffel symbols of the metric $g_{\mu\nu}$. The latter is defined by ([6], eq.(3.2.7))
$$g_{\mu\nu}={\d\xi^\al\over\d x^\mu}\eta_{\al\beta}
{\d\xi^\beta\over\d x^\nu}.\eqno(2.7)$$

Since the local inertial coordinates $\xi^\al$ can be changed by arbitrary Lorentz transformations, we can choose a Lorentz boost such that the two coordinate systems are at rest with respect to each other. Then we have
$${\d\xi^0\over\d x^j}=0,\quad {\rm and}\quad {\d\xi^j\over\d x^0}=0\eqno(2.8)$$
for $j=1,2,3$. This leads to
$$g_{00}=\B({\d\xi^0\over\d x^0}\B)^2,\quad g_{jk}=-\sum_{i=1}^3{\d\xi^i\over\d x^j}{\d\xi^i\over\d x^k}\eqno(2.9)$$
and $g_{0j}=0$. Now it follows
$$u^0={d\xi^0\over d\tau}={\d\xi^0\over\d x^0}{dx^0\over d\tau}=\sqrt{g_{00}}{dx^0\over d\tau}.\eqno(2.10)$$

In the following we assume that the non-diagonal elements $g_{jk}$ vanish. Then from the invariant
$$u^\al\eta_{\al\beta}u^\beta=(u^0)^2-\vec u^2=1={\d\xi^\al\over\d x^\mu}\eta_{\al\beta}
{\d\xi^\beta\over\d x^\nu}{dx^\mu\over d\tau}{dx^\nu\over d\tau}=$$
$$=g_{\mu\nu}{dx^\mu\over d\tau}{dx^\nu\over d\tau}=g_{00}\B({dx^0\over d\tau}\B)^2+g_{jj}\B({dx^j\over d\tau}\B)^2$$
we find
$$\vec V^2=-{1\over (u^0)^2}g_{jj}\B({dx^j\over d\tau}\B)^2=-\sum_{j=1}^3{g_{jj}\over g_{00}}{(dx^j/d\tau)^2\over (dx^0/d\tau)^2}
\eqno(2.11)$$
where (2.10) is used.

We want to specialize this for circular motion $r=const$. Note that in this case there is no first order Doppler effect for an observer in the center of the galaxy because $V_{\rm rel}=0$. But an observer on earth has $V_{\rm rel}\ne 0$ if he sees a spiral galaxy under some angle of inclination $i\ne 0$, for example.
Now the geodesic equation (2.6) for $\mu=1$ reads
$${d^2r\over d\tau^2}+\Gamma_{00}^1\B({dt\over d\tau}\B)^2+\Gamma_{11}^1\B({dr\over d\tau}\B)^2
+\Gamma_{22}^1\B({d\te\over d\tau}\B)^2+\Gamma_{33}^1\B({d\phi\over d\tau}\B)^2=0.\eqno(2.12)$$
Taking the circular orbit $r=const$ in the equatorial plane $\te=\pi/2$, then (2.12) gets simplified to
$$\B({d\phi\over d\tau}\B)^2=-{\Gamma_{00}^1\over\Gamma_{33}^1}\B({dt\over d\tau}\B)^2\eqno(2.13)$$
and from (2.11) we finally obtain the important relation
$$\vec V^2={g_{33}\over g_{00}}{\Gamma_{00}^1\over\Gamma_{33}^1}.
\eqno(2.14)$$

In the following we consider static spherically symmetric metrics of the form
$$ds^2=g_{\mu\nu}dx^\mu dx^\nu=e^a dt^2-e^bdr^2-r^2e^c(d\te^2+\sin^2\te d\phi^2),\eqno(2.15)$$
where $a(r), b(r), c(r)$ are functions of $r$ only. These metric functions are solutions of the gravitational field equations (Einstein's equation or the tensor 4-scalars theory). It is well known that the field equations do not determine the $g_{\mu\nu}$ uniquely, there is freedom to choose a gauge. In the standard gauge one puts $c=0$. It is argued that this can be achieved by redefining the radial variable as
$$r'=re^{c/2}.\eqno(2.16)$$
However, this new $r'$ is no longer the observable radial distance as defined above. Furthermore, $c(r)$ as solution of the field equations is not a priori known. Consequently, there is no clear definition of $r'$. Even if $c(r)$ were known, it would be different for different galaxies, so that $r'$ would be different; a very unpleasant situation.

The strongest argument against the standard gauge $c=0$ comes from observations. The vacuum solution with $c=0$ is unique (Birkhoff's theorem) and from (2.14) one finds $V^2\sim 1/r$ for large $r$. This does not fit the measured rotation curves for galaxies which show $V\to V_{\rm flat}=const$. The standard way out is to abandon the vacuum equations and assume some hypothetical dark matter. As long as this dark matter is not convincingly recorded one should study the other possibility of retaining $c(r)\ne 0$. Then the vacuum solution is no longer unique (see Sect.2 below). To fix it uniquely {\it we take the expression (2.14) for the circular velocity $V(r)$ as our non-standard gauge condition}. In other words, (i) every circular velocity $V(r)$ defines a particular gauge, (ii) since the circular velocity is an observable it is gauge invariant in contrast to the metric, therefore, our gauge fixing has a direct physical meaning. However, that means $V(r)$ must be given, so that the theory seems to have less predictive power. But what seems to be a weakness is a strength: $V(r)$ cannot be predicted on the basis of the vacuum equations alone, the dynamics of the normal matter must necessarily be taken into account. Indeed a uniform asymptotic velocity profile seems not to exist in reality. There is the baryonic Tully-Fisher relation [7-8] which relates the asymptotic velocity $V_{\rm flat}$ to the fourth root of the total baryonic mass $M^{1/4}$. But beside this the velocity curves at large $r$ have quite different behavior for different types of galaxies. Our gauge condition (2.14) precisely allows for this. The non-standard theory is much richer than standard general relativity where asymptotic flatness is usually assumed.

It has been argued that the function $c(r)$ specifies the relation between the radius $r$ and the circumference of circles, and therefore $c(r)$ should be zero. This geometric interpretation of $c(r)$ can be used for small bodies where both $r$ and the circumference can be measured by comparison with a measuring rod. But in a big galaxy one cannot do so. Then the (gauge-dependent) metric is only accessible by observing gravitational effects as the circular velocity (2.14). An additional geometric interpretation has no physical justification. We share Weinberg's non-geometric views ([6], sect.6/9):

``The geometric interpretation of the theory of gravitation has dwindled to a mere analogy, which lingers in our language in terms like 'metric', 'affine connection', and 'curvature', but is not otherwise very useful. The important thing is to be able to make predictions about images on the astronomers' photographic plates, frequencies of spectral lines, and so on,...''

The physical correctness or failure of our theory can only be decided by working out its further consequences (and by observing dark matter particles of course). Concerning this we refer to part III of this series which is now available in the arXiv 1205.4309.

\section{Spherically symmetric vacuum solutions}

According to [1] the vacuum field equations of the tensor-four-scalars theory are the modified Einstein equations 
$$R_{\mu\nu}-{1\over 2}g_{\mu\nu}R=
{16\pi G\over c^3}\Bigl\{-\d_\mu v_n\d_\nu v^n
+{1\over 2}g_{\mu\nu}g^{\al\beta}\d_\al v_n\d_\beta v^n\Bigl\}\eqno(3.1)$$
and the wave equation in the metric $g^{\al\beta}$
$$
\d_\al(\sqrt{-g}g^{\al\beta}\d_\beta v_n)=0.\eqno(3.2)$$
Here the index $n$ is raised and lowered with the Minkowski metric, but all other Greek indices with $g_{\mu\nu}$ as in ordinary general relativity.
$G$ is Newton's constant.

To study static spherically symmetric solutions we choose spherical coordinates
$$x^0=t,\quad x^1=r,\quad x^2=\te,\quad x^3=\phi\eqno(3.3)$$
and set the speed of light $c=1$. We assume the metric to be of the following non-standard form (2.12)
$$g_{00}=e^a,\quad g_{11}=-e^b,\quad g_{22}=-r^2e^c,\quad g_{33}=\sin^2\te g_{22}\eqno(3.4)$$
and zero otherwise. Since we look for static solutions we neglect the time derivatives $\d_0$ in the wave operator (3.2):
$$\d_\al (\sqrt{-g}g^{\al\beta}\d_\beta))=-\sin\te\d_r\Bigl(e^{d/2-b}r^2\d_r\Bigl)
-e^{d/2-c}\d_\te(\sin\te\d_\te)
-{e^{d/2-c}\over\sin\te}\d^2_\phi\eqno(3.5)$$
where $d=a+b+2c$. Then the wave equation for a static $v^n$ becomes
$${\d\over\d r}\Bigl(e^{d/2-b}r^2{\d\over\d r}\Bigl)v^n=e^{d/2-c}L^2v^n,
\eqno(3.6)$$
where $L^2$ is the quantum mechanical angular momentum operator squared. Since we restrict ourselves to spherically symmetric solutions the right-hand side vanishes.
Then only the time component $v^n=(v_0(r),0,0,0)$ can be different from 0, otherwise we would get non-diagonal elements in the metric.
The equation (3.6) can now be integrated once
$$v'_0(r)={A\over r^2}e^{b-d/2},\eqno(3.7)$$
where $A$ is an integration constant. The prime always means partial derivative with respect to $r$.

The Christoffel symbols and the Ricci tensor for the metric (3.4) have been computed in [1]. Only the diagonal elements $R_{\mu\mu}$ are different from zero.
Using (3.7) the modified Einstein equations (3.1) for the 00, 11, 22  and also 33 elements of the Ricci tensor give the following three differential equations
$$c''=-{3\over 4}c'^2+{1\over 2}b'c'+{1\over r}(b'-3c')+{1\over r^2}(e^{b-c}-1)
+{\al\over 2r^4}e^{2b-d}\eqno(3.8)$$
$$0={1\over 2}a'c'+{1\over r}(a'+c')+{c'^2\over 4}+{1\over r^2}\Bigl(1-e^{b-c}\Bigl)
+{\al\over 2r^4}e^{2b-d}\eqno(3.9)$$
$$a''+c''={1\over r}(b'-a'-2c')-{1\over 2}(a'^2-a'b'+a'c'-b'c'+c'^2)
+{\al\over r^4}e^{2b-d},\eqno(3.10)$$
where
$$\al=16\pi GA^2\eqno(3.11)$$
is another form of the constant of integration for $v'_0(r)$. In the following it is convenient to subtract (3.8) from (3.10)
$$a''={c'^2\over 4}-{a'\over 2}(a'-b'+c')+{1\over r}(c'-a')+{1\over r^2}(1-e^{b-c})$$
$$+{\al\over 2r^4}e^{2b-d}\eqno(3.12)$$
and to use this equation instead of (3.10). 

It is not hard to eliminate the function $b(r)$ from the equations. We solve eq.(3.9) for $b(r)$
$$b=c+2\log r-\log\Bigl(1-{\al\over 2r^2}e^{-a-c}\Bigl)+$$
$$+\log\Bigl({a'c'\over 2}+{a'+c'\over r}+{c'^2\over 4}+{1\over r^2}\Bigl).\eqno(3.13)$$
This enables us to eliminate $b$ in the remaining two equations (3.8) and (3.12). As we have already mentioned in [1] there is a degeneracy in these equations. Since this is an important property we give some details of the elimination process. Differentiating (3.13) we get
$$b'=c'+{2\over r}+\Bigl(1-{2r^2\over 2r^2-\al\exp(-a-c)}\Bigl)\Bigl(a'+c'+
{2\over r}\Bigl)+$$
$$+{1\over N}\Bigl(a''(2r^2c'+4r)+c''(2r^2a'+4r+2r^2c')-4(a'+c'+{2\over r})
\Bigl),\eqno(3.14)$$
with
$$N=2r^2a'c'+4r(a'+c')+r^2c'^2+4.\eqno(3.15)$$
We substitute this into (3.12) and solve for $c''$:
$$c''={a''\over a'+c'+2/r}\Bigl(c'+{2\over r}+{4c'\over ra'}+{c'^2\over a'}+{4\over r^2a'}\Bigl)-{N\over 2r^2-\al\exp(-a-c)}-{2\over r^2}.\eqno(3.16)$$
This enables us to simplify (3.14) considerably:
$$b'=2{a''\over a'}+a'+2c'+{4\over r}.\eqno(3.17)$$
Next we turn to (3.8) and substitute (3.17) for $b'$. This yields
$$c''={a''\over a'}\Bigl(c'+{2\over r}\Bigl)+{2\over r^2}+{2r^2a'c'+4r(a'+c')+r^2c'^2
+4\over 2r^2-\al\exp(-a-c)}.\eqno(3.18)$$
This is identical with (3.16) which shows the degeneracy in the system of vacuum field equations.
Summing up from (3.18) we find a class of solutions with one  free function $a(r)$ or
$c(r)$. This reflects the freedom of gauge.

\section{Integration of the vacuum equations}

It is convenient to simplify the equations by introducing the new metric function
$$f(r)=c(r)+2\log{r\over r_c}\eqno(4.1)$$
for $c(r)$. The quantity $r_c$ has been included for dimensional reason. Then eq.(3.18) assumes the simple form
$${f''\over f'}-{a''\over a'}=(a'+{f'\over 2})\Bigl(1-{\al\over r_c}e^{-a-f}\Bigl)^{-1},
\eqno(4.2)$$
and the circular velocity squared (2.11) simply becomes
$$V^2(r)={a'\over f'}\equiv u.\eqno(4.3)$$
There should be no confusion with the 4-velocity $u^\al$ in (2.3) because this quantity only appear in Sect.2.
We also introduce the new variable
$$w={a+c\over 2},\eqno(4.4)$$
so that
$$a'={2u\over r}{1+rw'\over 1+u}$$
$$c'={2\over r}{rw'-u\over 1+u}.\eqno(4.5)$$
Then (3.18) becomes
$${u'\over u}+{2r\over 2r^2-\al\exp(-2w)}{1+2u\over 1+u}(1+rw')=0.$$ 
After multiplication with $(1+u)/(1+2u)$ this is integrable in the form
$${d\over dr}\Bigl[\log{u\over\sqrt{1+2u}}+{1\over 2}\log\Bigl(r^2e^{2w}-{\al\over 2}
\Bigl)\Bigl]=0.\eqno(4.6)$$
Hence we get the integral
$${u^2(r^2e^{2w}-\al/2)\over 1+2u}={u^2\over 1+2u}\Bigl(r^2e^{a+c}-{\al\over 2}
\Bigl)=\beta,\eqno(4.7)$$
$\beta$ is a new constant of integration. 

Next we have to incorporate the circular velocity (4.3) which  leads to
$$a'=u\B(c'+{2\over r}\B).\eqno(4.8)$$
On the other hand from (4.7) we get
$$e^{a+c}={1\over r^2}\Bigl(\beta{1+2u\over u^2}+{\al\over 2}\Bigl).\eqno(4.9)$$
Taking the logarithm and differentiating we obtain $a'+c'$. Then using (4.8) we separately calculate
$$a'=-{2u'\over 1+2u+\gamma u^2}\eqno(4.10)$$
$$c'=-{2u'\over u(1+2u+\gamma u^2)}-{2\over r},\eqno(4.11)$$
with
$$\gamma={\al\over 2\beta}.\eqno(4.12)$$
Now everything is reduced to simple quadratures. The solution depending upon the existence of real roots of $1+2u+\gamma u^2$, we set, for $\gamma<1$,
	$$\gamma_1=1+\sqrt{1-\gamma},\quad \gamma_2=1-\sqrt{1-\gamma}.\eqno(4.13)$$
Then that we arrive at
	$$a = \left\{
		\begin{tabular}{l l}
			${1\over \sqrt{1-\gamma}}\log\B\vert{1+\gamma_2u\over 1+\gamma_1u}\B\vert+\log K_a$ & $\quad {\rm for}\quad\gamma<1$ \\
			$-{2 u\over u+1}+\log K_a$ & $\quad {\rm for}\quad\gamma=1$ \\
			$-{2\over\sqrt{\gamma-1}}\arctan{\sqrt{\gamma-1}u\over{1+u}}+\log K_a$ & $\quad {\rm for}\quad\gamma>1$ \\
		\end{tabular}
	\right.\eqno(4.14)$$
with $K_a$ an integration constant. Here we have chosen the indefinite integral in such a way that it is continuous over $0<u<\infty$ and $-\infty<\gamma<\infty$. In particular, near low velocities (remember that $u(r)$ is the circular velocity squared (4.3))
	$$a=-2u\B[1-u+{{4-\gamma}\over 3}u^2+(\gamma-2)u^3+O(u^4)\B]+\log K_a\qquad\forall\gamma,$$
whereas near $\gamma=1$
	$$a=-{2u\over 1+u}\B[1+{1\over 3}\B({u\over 1+u}\B)^2(1-\gamma)+{1\over 5}\B({u\over 1+u}\B)^4(1-\gamma)^2+O((1-\gamma)^3)\B]+\log K_a,$$
and near $\gamma=0$, which is the special case of ordinary general relativity,
	$$a=-\log(1+2u)+\B[{u(1+u)\over{1+2u}}-{1\over 2}\log(1+2u)\B]\gamma+O(\gamma^2)+\log K_a.$$
Function $c(r)$ obtains immediately from $a(r)$ using (4.12)
	$$c=-a+\log{\vert{1+2u+\gamma u^2}\vert\over{r^2u^2}}+\log{K_a K_c},\eqno(4.15)$$
with $K_c$ another integration constant. Substituting these results into (4.9) we get the following relation between the integration constants:
	$$K_aK_c=\beta\sgn (1+2u+\gamma u^2).\eqno(4.16)$$
Finally $b(r)$ also follows from $a$ by (3.13), (4.7) and (4.15)
	$$b=-a+\log{\B({u'\over u^2}\B)^2}+\log{K_a K_c}\eqno(4.17)$$

Taking the exponential we get the $00$-element of the metric
	$$e^a = \left\{
		\begin{tabular}{l l}
			$K_a \B\vert{1+\gamma_2u\over 1+\gamma_1u}\B\vert^{1/\sqrt{1-\gamma}}$ & $\quad {\rm for}\quad\gamma<1$ \\
			$K_a e^{-{2 u\over u+1}}$ & $\quad {\rm for}\quad\gamma=1$ \\
			$K_a e^{-{2\over\sqrt{\gamma-1}}\arctan{\sqrt{\gamma-1}u\over{1+u}}}$ & $\quad {\rm for}\quad\gamma>1$ \\
		\end{tabular}
	\right.\eqno(4.18)$$
In the same way we find from (4.15)
	$$e^c = \left\{
		\begin{tabular}{l l}
			$K_c {\vert 1+2u+\gamma u^2\vert\over{r^2 u^2}}\B\vert{1+\gamma_1u\over{1+\gamma_2u}}\B\vert^{1/\sqrt{1-\gamma}}$ & $\quad {\rm for}\quad\gamma<1$ \\
			$K_c {\vert 1+2u+\gamma u^2\vert\over{r^2 u^2}}e^{-{2 u\over u+1}}$ & $\quad {\rm for}\quad\gamma=1$ \\
			$K_c {\vert 1+2u+\gamma u^2\vert\over{r^2 u^2}}e^{-{2\over\sqrt{\gamma-1}}\arctan{\sqrt{\gamma-1}u\over{1+u}}}$ & $\quad {\rm for}\quad\gamma>1$ \\
		\end{tabular}
	\right.\eqno(4.19)$$
and from (4.17)
	$$e^b = \left\{
		\begin{tabular}{l l}
			$K_c \B({u'\over u^2}\B)^2 \B\vert{1+\gamma_1u\over 1+\gamma_2u}\B\vert^{1/\sqrt{1-\gamma}}$ & $\quad {\rm for}\quad\gamma<1$ \\
			$K_c \B({u'\over u^2}\B)^2 e^{2 u\over u+1}$ & $\quad {\rm for}\quad\gamma=1$ \\
			$K_c \B({u'\over u^2}\B)^2 e^{2\over\sqrt{\gamma-1}}\arctan{\sqrt{\gamma-1}u\over{1+u}}$ & $\quad {\rm for}\quad\gamma>1$ \\
		\end{tabular}
	\right.\eqno(4.20)$$

As a check, we can specialize these results to the classical Schwarzschild case which must be contained, of course. To this end, we first specialize our result to the pure tensor theory of general relativity, i.e. to the case $\alpha=0$. Then $\gamma=0$ by (4.12), and $\gamma_1=2$ and $\gamma_2=0$ by (4.13), so that (4.18-20) becomes
	$$e^a={K_a\over 1+2u},\quad e^b=K_c\B({u'\over u^2}\B)^2(1+2u),\quad e^c=K_c{(1+2u)^2\over r^2u^2}.\eqno(4.21)$$
Next we assume the Schwarzschild metric in standard form, as given eg. by [6] (8.2.12). By (2.14) the circular velocity squared is then equal to
	$$V^2=u={r_s\over 2(r-r_s)}\eqno(4.22)$$
with $r_s\equiv2GM$ the Schwarzschild radius. Inserting (4.22) in (4.21) yields
	$$e^a=K_a{r-r_s\over r},\quad e^b={4 K_c\over r_s^2}{r\over r-r_s},\quad e^c={4 K_c\over r_s^2}.\eqno(4.23)$$
After identifying
	$$K^a=1,\quad K_c={r_s^2\over 4}\eqno(4.24)$$
we recover the Schwarzschild metric and the standard gauge $c=0$ :
	$$e^a={r-r_s\over r},\quad  e^b={r\over r-r_s},\quad e^c=1,\eqno(4.25)$$
But clearly it would be dogmatic to forbid non-standard gauges. Moreover, in order to match the outer vacuum solution to an inner solution with normal matter, general relativity (4.21) might not be flexible enough. To be on the safe side we therefore continue to investigate the tensor-four-scalars theory.

\section{Fitting the tail of the rotation curve of M33}

Since we have neglected baryonic matter until now, we can compare the vacuum solution only with the tail of a real rotation curve where the contribution of the visible matter is small compared to the dark. Probably the best object for such a comparison is M33. This is a low luminosity spiral in the Local Group, it is dark matter dominated and near to us. There are fairly good data available which have been analyzed by Corbelli [8]. We are indebted to Edvige Corbelli for communicating the numerical values of Fig.5 of his paper.

As we have shown here the rotation curve $V(r)$ is not restricted by the vacuum solution. However our experience with power series expansions in [1] suggests that for large $r$ a series in powers of $r^{-1}$ should be a good representation of
$$V^2(r)=V_{\rm flat}^2+{V_1\over r}+{V_2\over r^2}+\ldots =u(r)\eqno(5.1)$$
According to the analysis of Corbelli the visible and dark contributions to $V^2(r)$ are equal around 7 kpc. Therefore, we have taken the last 8 points with $r>$ 7 kpc of the data and fitted them with weights given by $\sigma$ to four terms of (5.1). As the figure shows we obtain a very good fit with a reduced $\chi^2=0.1868$. This well passes a $\chi^2$ - test with $99\%$ confidence level and $8-4=4$ degrees of freedom. Of course this is not a test of the theory. A real test can only be carried out if additional data beside the rotation curve are available, for example lensing data. We have discussed such a test in [5].

Finally we note a peculiarity of the metric corresponding to the asymptotic behavior (5.1). From (4.16) and (4.18) we find the asymptotics of the metric functions
$$-g_{11}=e^b={L_1\over r^4}+O(r^{-5})\eqno(5.2)$$
$$-g_{22}=r^2e^c=L_2+O(r^{-1})\eqno(5.3)$$
and similarly for $g_{33}$. This gives for the determinant of the spatial part
$$-D=-g_{11}g_{22}g_{33}=e^{b+2c}r^4\sin^2\te={L_1L_2^2\over r^4}\sin\te+O(r^ {-5}).\eqno(5.4)$$
Since the volume measure is proportional to $\sqrt{-D}$ it follows that $\sqrt{-D}\sim r^{-2}$ is integrable at infinity, so that the metric gives a finite volume to the 3-space $(r,\te,\phi)$. That means the metric can only be locally true in a galaxy and its near neighborhood. In the large $V^2(r)$ must deviate from (5.1). Regarding M33 Corbelli has found that the ``dark halo'' extends to a distance comparable with the separation distance between M33 and its bright companion M31, the Andromeda galaxy. The model of one isolated galaxy in infinite space seems to be an unrealistic idealization. Therefore, it is no harm that we have no theory for this at present.

{\bf Acknowledgment.} We thank the referee for his criticism. It has enabled us to improve the manuscript considerably.

{\bf Figure caption:} Figure: M33 rotation curve (points) with a fit to eq.(5.1) for $R>7$ kpc.

\end{document}